\documentclass[10pt,twoside]{article}
\usepackage[latin1]{inputenc}
\usepackage[dvips]{graphicx}
\usepackage{amssymb}
\usepackage{amsmath}

\newtheorem{theorem}{Theorem}

\newtheorem{example}[theorem]{Example}

\begin{document}

\title{On the transformation of torques between the laboratory and center of
mass reference frames}
\author{Rodolfo A. Diaz\thanks{%
radiazs@unal.edu.co}, William J. Herrera\thanks{%
jherreraw@unal.edu.co} \\
Universidad Nacional de Colombia, \\
Departamento de Física. Bogotá, Colombia.}
\date{}
\maketitle

\vspace{-6mm}

\begin{abstract}
It is commonly stated in Newtonian Mechanics that the torque with respect to
the laboratory frame is equal to the torque with respect to the center of
mass frame plus a $\mathbf{R}\times \mathbf{F}$ factor, with $\mathbf{R}$
being the position of the center of mass and $\mathbf{F}$ denoting the total
external force. Although this assertion is true, there is a subtlety in the
demonstration that is overlooked in the textbooks. In addition, it is
necessary to clarify that if the reference frame attached to the center of
mass rotates with respect to certain inertial frame, the assertion is not
true any more.

PACS \{01.30.Pp, 01.55.+b, 45.20.Dd\}

\end{abstract}

In Newtonian Mechanics, we define the total external torque of a system of $%
n $ particles (with respect to the laboratory frame that we shall assume as
an inertial reference frame) as

\begin{equation*}
\mathbf{N}_{ext}=\sum_{i=1}^{n}\mathbf{r}_{i}\times \mathbf{F}_{i(e)}\ ,
\end{equation*}%
where $\mathbf{r}_{i},\mathbf{F}_{i\left( e\right) }$ denote the position
and total external force for the $i-th$ particle. The relation between the
position coordinates between the laboratory (L) and center of mass (CM)
reference frames \footnote{%
For the \emph{center of mass reference frame} we mean a system of reference
whose origin lies at the CM, at that is in relative translation (but not in
relative rotation) with respect to certain inertial frame.} is given by%
\begin{equation*}
\mathbf{r}_{i}=\mathbf{r}_{i}^{\prime }+\mathbf{R\ ,}
\end{equation*}%
with $\mathbf{R}$ denoting the position of the CM about the L, and $\mathbf{r%
}_{i}^{\prime }\ $denoting the position of the $i-$th particle with respect
to the CM, in general the prime notation denotes variables measured with
respect to the CM. An standard demonstration shows that \cite{mecanica}%
\begin{equation}
\mathbf{N}_{ext}=\left( \sum_{i=1}^{n}\mathbf{r}_{i}^{\prime }\times \mathbf{%
F}_{i\left( e\right) }\right) +\mathbf{R\times F\ ,}  \label{torqLCM1}
\end{equation}%
where $\mathbf{F}$ corresponds to the total external force on the system of particles
(measured by the L). It is usually said that the first term on the right
side of Eq. (\ref{torqLCM1}) provides the external torque relative to the
CM. Strictly speaking this is not the case, since $\mathbf{F}_{i(e)}$ is a
force measured with respect to the L system, and since the CM is not in
general an inertial reference frame, the force measured by the CM is not
equal to the force in the L system. As it is well known from the theory of
non-inertial systems \cite{mecanica2}, the total force on the $i-$th
particle measured about the CM reads%
\begin{equation}
\mathbf{F}_{i}^{\prime }=\mathbf{F}_{i}-m_{i}\mathbf{A}_{CM}\ .
\label{F total}
\end{equation}%

Where $\mathbf{A}_{CM}$ denotes the acceleration of the CM with respect to the L. Taking into account that the force on the $i-$th particle is given by the
sum of the external forces plus the internal ones we have%
\begin{equation}
\mathbf{F}_{i(e)}^{\prime }+\sum_{k=1}^{n}\mathbf{F}_{ik}^{\prime }=\mathbf{F%
}_{i(e)}+\sum_{k=1}^{n}\mathbf{F}_{ik}-m_{i}\mathbf{A}_{CM}\ ,
\label{F total2}
\end{equation}%
where $\mathbf{F}_{ik}$ denotes the internal force on the $i-$th particle
due to the $k-$th particle, and $m_{i}$ is the corresponding mass. Now, if we take
into account that the internal forces are independent of the reference frame%
\footnote{%
In the case of central internal forces it is clear since they are functions
of the relative positions among different pairs of particles. Nevertheless,
even in the case in which the forces are not central, they depend on
relative positions, relative velocities, relative acelerations etc. So that
the invariance under different frames still holds.}, and using Eq. (\ref{F total2}) we get%
\begin{equation}
\mathbf{F}_{i(e)}^{\prime }=\mathbf{F}_{i(e)}-m_{i}\mathbf{A}_{CM}\ .
\label{external}
\end{equation}

From Eq. (\ref{external}), the external torque about the CM becomes%
\begin{eqnarray*}
\mathbf{N}_{CM} &=&\sum_{i=1}^{n}\mathbf{r}_{i}^{\prime }\times \mathbf{F}%
_{i(e)}^{\prime }=\sum_{i=1}^{n}\mathbf{r}_{i}^{\prime }\times \left[ 
\mathbf{F}_{i(e)}-m_{i}\mathbf{A}_{CM}\right] \\
&=&\sum_{i=1}^{n}\mathbf{r}_{i}^{\prime }\times \mathbf{F}_{i\left( e\right)
}-M\left( \frac{1}{M}\sum_{i=1}^{n}m_{i}\mathbf{r}_{i}^{\prime }\right)
\times \mathbf{A}_{CM}\ ,
\end{eqnarray*}%
the term in parenthesis corresponds to the position of the CM with respect
to the CM itself, therefore it clearly vanishes, from which we see that 
\begin{equation}
\sum_{i=1}^{n}\mathbf{r}_{i}^{\prime }\times \mathbf{F}_{i\left( e\right)
}=\sum_{i=1}^{n}\mathbf{r}_{i}^{\prime }\times \mathbf{F}_{i\left( e\right)
}^{\prime }=\mathbf{N}_{CM}\ ,  \label{torqLCM}
\end{equation}%
replacing Eq. (\ref{torqLCM}) into Eq. (\ref{torqLCM1}) we get%
\begin{equation}
\mathbf{N}_{ext}=\left( \sum_{i=1}^{n}\mathbf{r}_{i}^{\prime }\times \mathbf{%
F}_{i\left( ext\right) }^{\prime }\right) +\mathbf{R\times F\ .}
\label{torqLCM2}
\end{equation}%
From Eq. (\ref{torqLCM2}) we can assert that \emph{the total external torque
about the laboratory is equal to the external torque about the center of
mass plus the torque equivalent to a particle located at the position of the
center of mass undergoing the total external force of the system of
particles. }

This coincides with the assertion given in commontexts. However, such
statement follows from Eq. (\ref{torqLCM2}) and not from Eq. (\ref{torqLCM1}%
) as appears in the literature. Moreover, as it is clear from the
development above, the demonstration of Eq. (\ref{torqLCM}) is the clue to
asseverate this statement. In turn, Eq. (\ref{torqLCM}) is satisfied because
the fictitious forces do not contribute to the total external torque.

Finally, this clarification is also necessary to establish the equation%
\begin{equation*}
\frac{d\mathbf{L}_{CM}}{dt}=\mathbf{N}_{CM}\ ,
\end{equation*}%
with $\mathbf{L}_{CM}$ denoting the total angular momentum of the system of
particles about the CM. As it is well known, this equation is valid even if
the center of mass is a non-inertial system of reference \cite{mecanica}, in
whose case the fictitious forces should be included in the demonstration.

\begin{figure}[tbh]
\begin{center}
\includegraphics[height=3.5cm]{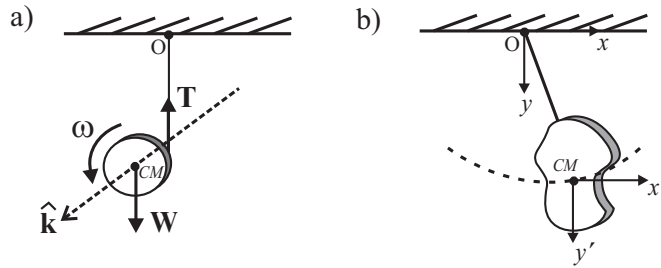}
\end{center}
\caption{(a) A yo-yo unwinding from a string. Its CM is accelerated with respect
to an inertial reference frame. (b) A physical pendulum hanging from a fixed point $O$. The system attached to $O$ is considered inertial, while the system associated with the CM is in relative translation with respect to the latter. In this case the acceleration of the CM is not constant, notwithstanding the fictitious force associated with the CM system does not give a torque.}
\label{fig:yoyo}
\end{figure}

\begin{example}
Let us consider the case of a yo-yo of mass $M$ and radius $b$ which falls unwinding by means of a string, as shown in Fig. \ref{fig:yoyo}a. In a system
of reference located at $O$, the forces over the yo-yo are the weight and
the tension, both of them produces a torque and the problem is quite
complicate. In the system of reference of the CM we have an additional
fictitious force. The weight has null torque, the tension gives a torque of $%
bT\mathbf{\hat{k}}$, $\mathbf{\ }$and the torque associated with the
fictitious force can be written as%
\begin{equation*}
\mathbf{N}_{fict}=-\int dm\ \mathbf{r}^{\prime }\times \mathbf{a}_{CM}\ .
\end{equation*}

Where $\mathbf{a}_{CM}$ denotes the acceleration of the CM with respect to $%
O $, factorizing $\mathbf{a}_{CM}$ we get $\mathbf{N}_{fict}=-M\mathbf{R'}%
\times \mathbf{a_{CM}}=0,$ and hence the fictitious torque vanishes. Therefore we
get the result usually found in textbooks%
\begin{equation*}
Mg-T-Ma=0\ ,\ \frac{d\mathbf{L}_{CM.}}{dt}=I\frac{d\omega }{dt}=bT\widehat{%
\mathbf{k}}.
\end{equation*}%
Although the result is the same as in the literature, the correct procedure
requires to take into account the fact that the system of reference attached
to the CM is non inertial.
\end{example}

\begin{figure}[tbh]
\begin{center}
\includegraphics[height=3.5cm]{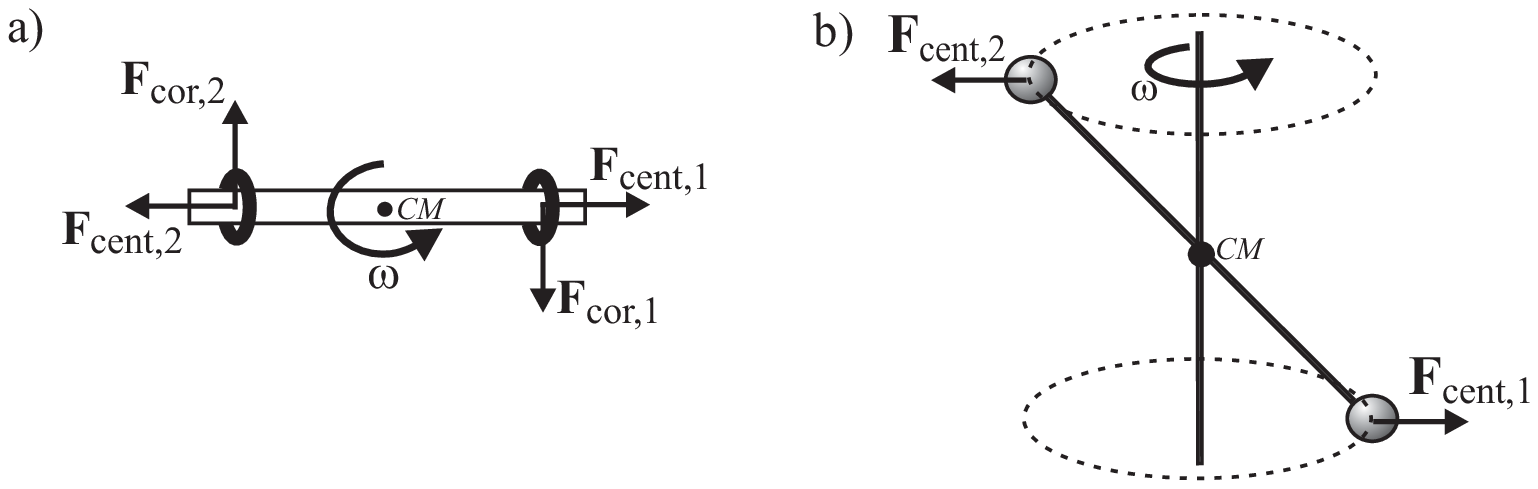}
\end{center}
\caption{(a) Aerial view of a couple of identical beads sliding on a
rotating wire. The $Z-$axis is out of the paper. The centrifugal and
coriolis forces are indicated in the diagram. (b) Two identical particles
joined by a thin rod of negligible mass, and rotating at constant angular
velocity $\protect\omega $, the centrifugal forces are indicated in the
diagram and no coriolis force appears in a system that rotates with the
particles. In both diagrams we only show the fictitious forces.}
\label{fig:rot}
\end{figure}

In the case in which the acceleration of the center of mass is constant, we
can relate the fact that the fictitious forces do not produce torque, with
the principle of equivalence. According to this principle, the resultant
fictitious force on the system of particles is equivalent to a weight owe to a uniform
gravitational field. Consequently, it appears as a gravity force acting on
the CM (see Sec. 8.4 in Ref. \cite{mecanica2}) and therefore the torque with
respect to the CM is clearly zero. Nevertheless, we point out that according
to our demostration, the torque due to the fictitious forces vanishes even
if the acceleration $\mathbf{A}_{CM}$ is not constant, in whose case the
equivalence principle does not hold. Fig. \ref{fig:rot}b shows a system in which the CM reference frame is in relative translation with respect to an inertial frame, in this case the torque associated with the fictitious force vanishes, though $\mathbf{A}_{CM}$ is not constant and the equivalence principle cannot be applied.

In the previous treatment we have implicitly assumed that the center of mass
reference frame defines coordinate axes that are only in relative
translation with respect to the inertial system. Now let us assume that the
center of mass is at rest with respect to a certain inertial frame, but the
axes of coordinates associated to the CM defines a rotating system with
angular velocity $\mathbf{\Omega }$ with respect to the axes of coordinates
of the inertial frame. As it is well known, the vectors of position coincide in both systems of reference (i.e. $\mathbf{r}_{i}^{\prime }=%
\mathbf{r}_{i}$)\ but not their time derivatives. The total torque on the
system is%
\begin{equation}
\mathbf{N}_{total}^{\prime }=\sum_{i=1}^{n}\mathbf{r}_{i}\times \mathbf{F}%
_{i}^{\prime }=\sum_{i=1}^{n}\mathbf{r}_{i}\times \left[ \sum_{i\neq k}^{n}%
\mathbf{F}_{ik}+\mathbf{F}_{i\left( e\right) }-2m_{i}\mathbf{\Omega }\times 
\mathbf{v}_{i}^{\prime }-m_{i}\mathbf{\Omega }\times \left( \mathbf{\Omega }%
\times \mathbf{r}_{i}\right) \right]
\end{equation}

where $\sum_{i\neq k}^{n}\mathbf{F}_{ik}+\mathbf{F}_{i\left( e\right) }$
defines the force on the $i-$th particle, measured by the inertial reference frame and $-2m_{i}%
\mathbf{\Omega }\times \mathbf{v}_{i}^{\prime }$, $-m_{i}\mathbf{\Omega }%
\times \left( \mathbf{\Omega }\times \mathbf{r}_{i}\right) \ $refers to the
coriolis and centrifugal forces respectively. For simplicity, we shall
assume that the internal forces are central and that they depend on relative
positions only (and not on relative velocities or higher time derivatives),
so that the torque associated with the internal forces vanishes, and only the
external torque contributes, yielding%
\begin{eqnarray*}
\mathbf{N}_{ext}^{\prime } &=&\sum_{i=1}^{n}\mathbf{r}_{i}\times \left[ 
\mathbf{F}_{i\left( e\right) }-2m_{i}\mathbf{\Omega }\times \mathbf{v}%
_{i}^{\prime }-m_{i}\mathbf{\Omega }\times \left( \mathbf{\Omega }\times 
\mathbf{r}_{i}\right) \right] \\
&=&\mathbf{N}_{ext}+\mathbf{N}_{fict}
\end{eqnarray*}%
with 
\begin{equation}
\mathbf{N}_{fict}=-\sum_{i=1}^{n}\left\{ m_{i}\mathbf{r}_{i}\times \left[ 2%
\mathbf{\Omega }\times \mathbf{v}_{i}^{\prime }+\mathbf{\Omega }\times
\left( \mathbf{\Omega }\times \mathbf{r}_{i}\right) \right] \right\}
\label{torqrot}
\end{equation}%
Eq. (\ref{torqrot}) shows that in the most general case, the fictitious
forces could produce a torque. Let us illustrate this issue with the
following example

\begin{example}
A couple of identical beads are sliding without friction on a rigid wire (of
negligible mass) rotating at constant angular speed $\omega $ (see Fig. \ref%
{fig:rot}a). In this case the CM of the system is at rest with respect to an
inertial frame. But we can define a rotating system with the origin in the
CM and rotating with the wire. The coriolis and centrifugal forces on each
bead are plotted in Fig. \ref{fig:rot}a. Since we are interested in the torque produced by fictitious forces
only, we do not consider the torque coming from the (real) normal force. Assuming the axis
of rotation to be along with the $z$ axis we obtain%
\begin{equation*}
\mathbf{N}_{fict}=\mathbf{r}_{1}\times \left( \mathbf{F}_{cor,1}+\mathbf{F}%
_{cent,1}\right) +\mathbf{r}_{2}\times \left( \mathbf{F}_{cor,2}+\mathbf{F}%
_{cent,2}\right) 
\end{equation*}%
the centrifugal forces are parallel to the corresponding vectors of position
so that their contribution vanish, we also see that $\mathbf{r}_{2}=-\mathbf{%
r}_{1}$ and $\mathbf{F}_{cor,2}=-\mathbf{F}_{cor,1}$ and the fictitious
torque becomes%
\begin{equation*}
\mathbf{N}_{fict}=2\mathbf{r}_{1}\times \mathbf{F}_{cor,1}=-2r_{1}\omega
v_{1}^{\prime }\widehat{\mathbf{k}}=-2r_{1}\dot{r}_{1}\omega \widehat{%
\mathbf{k}}
\end{equation*}%
the values of $r_{1},$ $\dot{r}_{1}$ can be calculated to be (see example
8.7 in Ref. \cite{mecanica2})%
\begin{equation*}
r_{1}=Ae^{\omega t}+Be^{-\omega t}\ \ ;\ \dot{r}_{1}=\omega \left(
Ae^{\omega t}-Be^{-\omega t}\right) 
\end{equation*}%
and the fictitious torques do not vanish. It is not difficult to find
examples in which the centrifugal force could also contribute to the
fictitious torque. The reader could try with the system described by Fig \ref%
{fig:rot}b.
\end{example}

In conclusion, we can asseverate that%
\begin{equation*}
\mathbf{N}_{ext}=\mathbf{N}_{CM}+\mathbf{R}\times \mathbf{F}_{ext}\ ;\ \frac{%
d\mathbf{L}_{CM}}{dt}=\mathbf{N}_{CM}
\end{equation*}%
only if the system of reference attached to the CM is not in relative
rotation with respect to certain inertial frame. It is related with the fact
that the fictitious forces produced by relative accelerated translation do
not produce torque, while the fictitious forces associated with relative
rotation do.

\begin{figure}[tbh]
\begin{center}
\includegraphics[height=3.5cm]{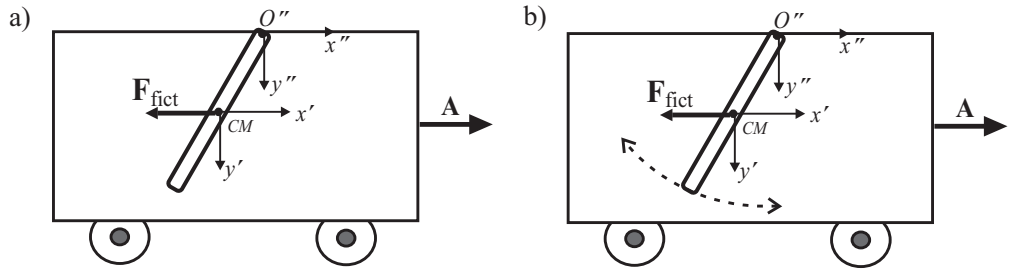}
\end{center}
\caption{In both Figs. the car has an acceleration $\mathbf{A}$ and we assume 2 dimensional reference frames for simplicity: (a) The thin stick is at rest with respect to the car. The acceleration of the CM of the thin stick is precisely $\mathbf{A}$ and the system of reference $S"$ has the same acceleration. However the origin of $S"$ does not lie at the CM of the stick. It can be shown that in this case the fictitious force produces a torque. (b) The stick is oscillating around its equilibrium position. In this case the acceleration of $S"$ does not coincide with the acceleration of the center of mass of the stick. Once again, there is a torque associated with the fictitious force.}
\label{fig:tren}
\end{figure}

A final comment about a more general framework, let us assume a non-inertial system (denoted by $S"$) in relative translation with respect to an inertial frame. If the acceleration of $S"$ coincides with $\mathbf{A}_{CM}$, but the origin of $S"$ is not located at the CM, the fictitious forces associated with $S"$ \textbf{do produce a torque}; Fig. \ref{fig:tren}a shows an example. In addition, if the acceleration of $S"$ is not the same as the acceleration of the CM, the fictitious forces associated with $S"$ also produces a torque as it is shown in Fig. \ref{fig:tren}b. These fictitious torques can be calculated by assuming that the resultant fictitious force is applied to the CM even if $\mathbf{A}_{CM}$ depends on the time.

\end{document}